\documentclass[english,aps,twocolumn,longbibliography]{revtex4-1}
\usepackage[T1]{fontenc}
\usepackage[latin9]{inputenc}
\usepackage{color}
\usepackage{float}
\usepackage{amstext}
\usepackage{amssymb}
\usepackage{graphicx}
\usepackage{esint}

\makeatletter

\providecommand{\tabularnewline}{\\}

\@ifundefined{textcolor}{}
{%
 \definecolor{BLACK}{gray}{0}
 \definecolor{WHITE}{gray}{1}
 \definecolor{RED}{rgb}{1,0,0}
 \definecolor{GREEN}{rgb}{0,1,0}
 \definecolor{BLUE}{rgb}{0,0,1}
 \definecolor{CYAN}{cmyk}{1,0,0,0}
 \definecolor{MAGENTA}{cmyk}{0,1,0,0}
 \definecolor{YELLOW}{cmyk}{0,0,1,0}
}

\makeatother

\usepackage{babel}
\begin{document}

\title{Quantum Theory of Continuum Optomechanics}

\author{Peter Rakich}

\address{Department of Applied Physics, Yale University, New Haven, Connecticut
06520, USA}

\author{Florian Marquardt}

\address{Max Planck Institute for the Science of Light, Staudtstr. 2, 91058
Erlangen, Germany}

\address{Department of Physics, University of Erlangen-Nuremberg, Staudtstr.
7, 91058 Erlangen, Germany}
\begin{abstract}
We present the basic ingredients of continuum optomechanics, i.e.
the suitable extension of cavity optomechanical concepts to the interaction
of photons and phonons in an extended waveguide. We introduce a real-space
picture and argue which coupling terms may arise in leading order
in the spatial derivatives. This picture allows us to discuss quantum
noise, dissipation, and the correct boundary conditions at the waveguide
entrance. The connections both to optomechanical arrays as well as
to the theory of Brillouin scattering in waveguides are highlighted.
We identify the 'strong coupling regime' of continuum optomechanics
that may be accessible in future experiments.
\end{abstract}
\maketitle

\section{Introduction}

Cavity optomechanics \cite{aspelmeyer_cavity_2014} is a very active
research area at the interface of nanophysics and quantum optics.
Its aim is to exploit radiation forces to couple optical and vibrational
modes in a confined geometry, with applications ranging from sensitive
measurements, wavelength conversion, and squeezing all the way to
fundamental questions of quantum physics. The paradigmatic cavity-optomechanical
system is zero-dimensional, i.e. there is no relevant notion of spatial
distance or dimensionality that would affect the dynamics in an essential
way.

However, even though the vast majority of optomechanical systems rely
on an optical cavity, there are a number of implementations that evade
this paradigm. In particular, optomechanical effects are observed
in waveguide-type structures, where both the optical field and the
vibrations propagate in 1D, with the potential to uncover new classical
and quantum phenomena. For example, these include waveguides fabricated
on a chip \cite{rakich_giant_2012,van_laer_interaction_2015} as well
as thin membranes suspended in hollow core fibres \cite{butsch_cw-pumped_2014}.
There have also been hybrid approaches, e.g., where the light propagates
along the waveguide but couples to a localized mechanical mode \cite{li_harnessing_2008},
or with acoustic waves in whispering-gallery microresonators \cite{bahl_observation_2012,bahl_brillouin_2013}.

Coupling light and sound inside a waveguide has long been the subject
of studies on Brillouin (and Raman) scattering in fibres \cite{shen_theory_1965,shelby_guided_1985,boyd_nonlinear_2013,agrawal_nonlinear_2012}.
This connection, between Brillouin physics and optomechanics, has
recently been recognized as potentially fertile, and during the past
year, first theoretical studies emphasizing this connection have emerged.
The cavity-optomechanical coupling in a torus has been derived by
starting from the known description of Brillouin interactions in an
infinitely extended waveguide \cite{van_laer_unifying_2016}. Conversely,
the Hamiltonian coupling light and sound in such waveguides has been
derived starting from the microscopic optomechanical interaction \cite{laude_lagrangian_2015,sipe_hamiltonian_2016,zoubi_optomechanical_2016,kharel_noise_2016},
including both boundary and photoelastic terms and fully incorporate
geometric and material properties of the system. These works represent
important bridges between the rapidly developing field of optomechanics
and the significantly more advanced field of Brillouin scattering. 

Independently, the role of dimensionality has also been emphasized
for several years now in another area of optomechanics: Discrete optomechanical
arrays, i.e. periodic (1D or 2D) lattices of coupled optical and vibrational
modes. These could be implemented in various settings, including photonic
crystals \cite{safavi-naeini_two-dimensional_2014}, coupled optical
disk resonators \cite{zhang_synchronization_2015}, or stacks of membranes.
Recent theoretical studies have revealed their interesting properties,
including the generation of photon-phonon bandstructures \cite{chang_slowing_2011,chen_photon_2014,schmidt_optomechanical_2015-1},
synchronization and nonlinear dynamics \cite{heinrich_collective_2011},
effects of long-range coupling \cite{xuereb_strong_2012,schmidt_optomechanical_2012,xuereb_reconfigurable_2014},
quantum many-body physics \cite{ludwig_quantum_2013}, and the creation
of artificial gauge fields and topological transport \cite{schmidt_optomechanical_2015,peano_topological_2015,walter_dynamical_2015}.

In the present manuscript, our aim is to establish simplified foundations
for ``continuum optomechanics'', i.e., optomechanics in 1D waveguides
without cavity modes: (i) We introduce a real-space picture and discuss
how one can enumerate the possible coupling terms to leading order
in spatial derivatives. (ii) We show how the continuum limit arises
starting from discrete optomechanical arrays, thereby connecting Brillouin
physics and these lattice structures. (iii) We include dissipation
and quantum noise, deriving the quantum Langevin equations and the
boundary conditions at the input of the waveguide. (iv) We identify
the 'strong coupling regime' that may be accessible in the future.
(v) We provide an overview of experimentally achieved coupling parameters.

\section{Continuum Optomechanics in a Real-Space Formulation}

The usual cavity-optomechanical interaction Hamiltonian connects the
photon field $\hat{a}$ of a localized optical mode and the phonon
field $\hat{b}$ of a localized vibrational mode. It is of the parametric
form \cite{aspelmeyer_cavity_2014}

\begin{equation}
-\hbar g_{0}\hat{a}^{\dagger}\hat{a}(\hat{b}+\hat{b}^{\dagger})\,.
\end{equation}
Our goal is to generalize this in the most straightforward way to
the case of 1D continuum fields. We will do so in in real space, using
phenomenological considerations. For an evaluation of the coupling
constants for particular geometries one would resort to microscopic
approaches, such as those presented recently in \cite{laude_lagrangian_2015,sipe_hamiltonian_2016,zoubi_optomechanical_2016,sarabalis_guided_2016}.
While these approaches are powerful, and necessary to design an experimental
system, they are involved and rather complex as an entry point into
continuum optomechanics. Therefore, a phenomenological analysis can
be useful in its own right. For many purposes, the level of detail
provided here will be sufficient \textendash{} similar to cavity optomechanics,
where the microscopic calculation of $g_{0}$ is left as a separate
task. Moreover, a real-space picture is particularly useful in spatially
inhomogeneous situations, such as those brought about by disorder,
design, or nonlinear structure formation.

\begin{figure}
\includegraphics[width=1\columnwidth]{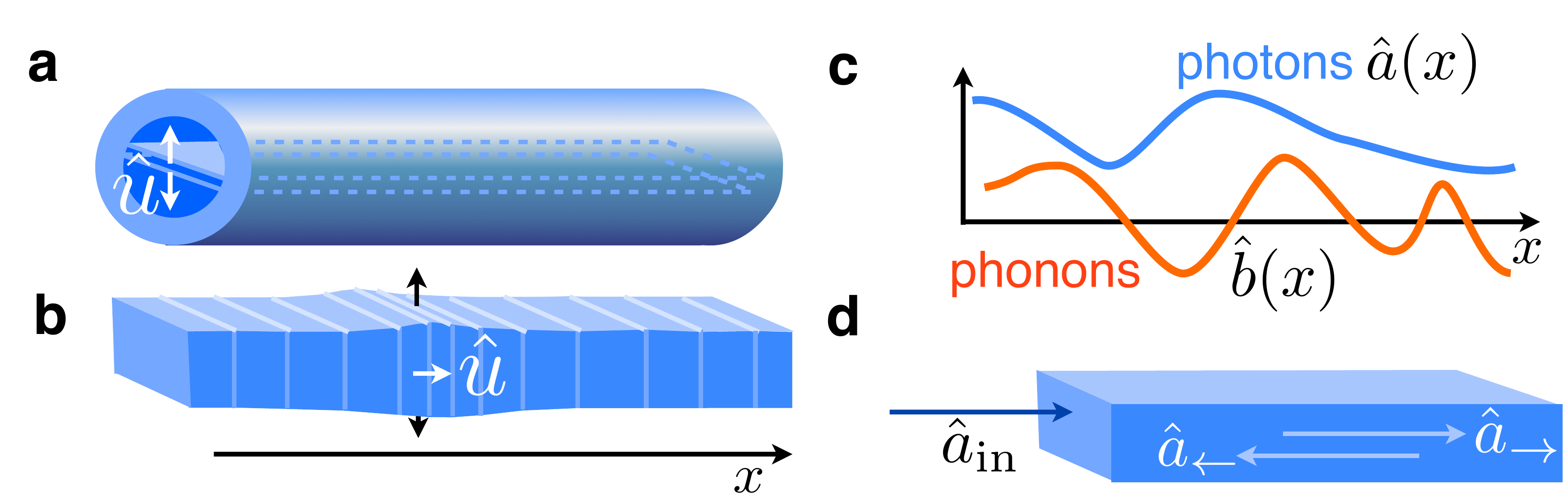}

\protect\caption{\label{FigBasicSetup}Continuum optomechanics. (a) Dual nanoweb structure
in photonic crystal fibres; displacement field $\hat{u}(x)$ describing
deflection of the membranes. (b) Nanobeam; $\hat{u}(x)$ describing
longitudinal displacement. (c) Photon and phonon fields in a 1D waveguide
geometry. (d) Situation at the input facet, relevant for the boundary
conditions.}

\end{figure}

We introduce photon and phonon fields $\hat{a}(x)$ and $\hat{b}(x)$,
respectively, for the waveguide geometry that we have in mind (Fig.~\ref{FigBasicSetup}).
In contrast to prior treatments, we do not assume the fields to be
sharply peaked around a particular wavevector \cite{laude_lagrangian_2015,sipe_hamiltonian_2016,zoubi_optomechanical_2016,kharel_noise_2016}.
This keeps our approach general and simplifies the representation
of the interacting fields, especially for situations with strongly
nonlinear dynamics. For example, this apprach avoids the need to treat
cascaded forward-scattering with an infinite number of photon fields
\cite{wolff_cascaded_2016}. The fields are normalized such that the
total photon number in the entire system would be $\int dx\,\hat{a}^{\dagger}(x)\hat{a}(x)$,
and likewise for the phonons. In addition, the fields obey the usual
bosonic commutation relations for a 1D field, e.g. $\left[\hat{a}(x),\hat{a}^{\dagger}(x')\right]=\delta(x-x')$.
For a nearly monochromatic wave packet of frequency $\omega$, the
energy density at position $x$ is $\hbar\omega\left\langle \hat{a}^{\dagger}(x)\hat{a}(x)\right\rangle $,
and the power can be obtained by multiplying this by the group velocity.
The plane-wave normal modes would be $\hat{a}(k)=\int\left(dx/\sqrt{2\pi}\right)\, e^{-ikx}\hat{a}(x)$,
with $\left[\hat{a}(k),\hat{a}^{\dagger}(k')\right]=\delta(k-k')$.

The normalized mechanical displacement field can be written as $\hat{u}(x)=\hat{b}(x)+\hat{b}^{\dagger}(x)$.
The physical displacement at any given point will be obtained by multiplying
with the mode function. As is well-known from standard cavity optomechanics,
any arbitrariness arising from the mode function normalization is
avoided by formulating everything in terms of $\hat{a}(x)$ and $\hat{b}(x)$,
since their normalization is directly tied to the overall energy in
the system.

The most obvious continuum optomechanical interaction can be written
down as a direct generalization of the cavity case:

\begin{equation}
\hat{H}_{{\rm int}}=-\hbar\tilde{g}_{0}\int dx\,\hat{a}^{\dagger}(x)\hat{a}(x)\hat{u}(x)\,.\label{InteractionSimpleVersion}
\end{equation}
Here $\tilde{g}_{0}$ defines the \emph{continuum} optomechanical
coupling constant, which replaces the usual single-photon cavity-optomechanical
coupling $g_{0}$. We note that $\tilde{g}_{0}$ has dimensions of
frequency times the square root of length. Its meaning can be understood
best in the following way: If there is a mechanical deflection $\left\langle \hat{b}\right\rangle =1/\sqrt{l}$,
corresponding to 1 phonon per length $l$, then the energy of any
photon is shifted by $-\hbar\tilde{g}_{0}2/\sqrt{l}$. We will comment
more on the $\sqrt{l}$ dependence when we make the connection to
discrete optomechanical arrays.

While Eq.~(\ref{InteractionSimpleVersion}) is a plausible ansatz,
it turns out to be only a part of the full interaction. Specifically,
in a real-space formulation of the continuum case, derivative terms
may appear, which we will now discuss.

There are both boundary and bulk terms that contribute to the shift
of optical frequency when a dielectric is deformed, as is well-known
for optomechanics and has also been discussed recently in the present
context \cite{sipe_hamiltonian_2016,zoubi_optomechanical_2016}. The
boundary terms are proportional to the displacement $\hat{u}$, and
as such their most natural representation is in the form of the ansatz
given above, if $\hat{u}$ is chosen to represent the deflection of
the boundary (more on this, see below). The bulk terms (photoelastic
response), however, depend only on the spatial derivatives of the
displacement field. In particular, this also involves derivatives
along the longitudinal (waveguide) direction, and these terms then
naturally lead to an expression

\begin{equation}
\hat{H}_{{\rm int}}^{++-}=-\hbar g_{0}^{++-}\int dx\,\hat{a}^{\dagger}(x)\hat{a}(x)\partial_{x}\hat{u}(x)\,.
\end{equation}
We have introduced a superscript $++-$ for the coupling constant,
indicating the possible presence of a derivative: $\partial_{x}\hat{u}$
changes sign if we set $\hat{u}(x)\mapsto\hat{u}(-x)$, so we associate
a negative signature.

It is important to note that the shape of the Hamiltonian depends
on the physical meaning of the displacement $\hat{u}$, which is to
some degree a matter of definition. We have to distinguish the full
vector field $\hat{\vec{u}}(\vec{r})$, which is defined unambiguously,
from the reduced one-dimensional field $\hat{u}(x)$ that forms the
object of our analysis. As a concrete example, consider longitudinal
waves on a nanobeam. The 1D field $\hat{u}(x)$ could then be defined
as the longitudinal displacement, evaluated at the beam center (see
Fig.~\ref{FigBasicSetup}b, white arrow). In that case the density
change, responsible for the photoelastic coupling, is proportional
to $\partial_{x}\hat{u}(x)$. At the same time, a finite Poisson ratio
will lead to a lateral expansion of the beam, i.e. a motion of the
surface. The surface deflection will be proportional to the density
change, and thus also determined by $\partial_{x}\hat{u}(x)$. However,
we could have defined $\hat{u}(x)$ differently, namely to represent
directly the surface deflection (Fig.~\ref{FigBasicSetup}b, black
arrows). In that case, the density change would be given by $\hat{u}(x)$.
Two different, equally valid definitions of $\hat{u}(x)$ would thus
lead to different expressions in the Hamiltonian.

Besides the appearance of derivatives $\partial_{x}\hat{u}$, we may
also encounter derivatives of the electric field. It is well-known
that electromagnetic waves inside matter can also have longitudinal
components, which change sign upon inversion of the propagation direction
(in contrast to the transverse fields). Consequently, the electromagnetic
mode functions depend on the direction of the wavevector, i.e. $\vec{E}(\vec{r})=\vec{E}_{k}(\vec{r}_{\perp})\exp(ikx)$.
Upon going to our reduced 1D real-space description, this dependence
on the sign of $k$ leads to terms that are the derivatives of the
1D field, since for a plane wave $\hat{a}(x)=\hat{a}e^{ikx}$ we have
$\partial_{x}\hat{a}(x)=ik\hat{a}(x)$ . Any terms in the full 3D
light-matter coupling that depended on the longitudinal components
(that change sign with $k$) will give rise to such derivatives of
the 1D fields. 

In summary, the possible combinations of derivatives that can occur
are listed in table \ref{PossibleCouplings}. 

\begin{table}[H]
\begin{centering}
\begin{tabular}{|c|c|}
\multicolumn{1}{c}{Even coupling terms} & \multicolumn{1}{c}{Odd coupling terms}\tabularnewline
\hline 
\hline 
$g_{0}^{+++}\hat{a}^{\dagger}\hat{a}\hat{u}$ & $g_{0}^{++-}\hat{a}^{\dagger}\hat{a}\left(\partial_{x}\hat{u}\right)$\tabularnewline
$g_{0}^{--+}\left(\partial_{x}\hat{a}^{\dagger}\right)\left(\partial_{x}\hat{a}\right)\hat{u}$ & $g_{0}^{-++}\left(\partial_{x}\hat{a}^{\dagger}\right)\hat{a}\hat{u}+{\rm h.c.}$\tabularnewline
$g_{0}^{-+-}(\partial_{x}\hat{a}^{\dagger})\hat{a}(\partial_{x}\hat{u})+{\rm h.c.}$ & $g_{0}^{---}\left(\partial_{x}\hat{a}^{\dagger}\right)\left(\partial_{x}\hat{a}\right)\left(\partial_{x}\hat{u}\right)$\tabularnewline
\hline 
\end{tabular}
\par\end{centering}

\protect\caption{\label{PossibleCouplings}Possible coupling terms (to leading order
in the derivatives) for continuum optomechanics, formulated in real-space.
The Hamiltonian is of the form $-\hbar\int dx(...)$, with the integrand
$(...)$ containing one or more terms displayed here.}

\end{table}

This is a complete list of the coupling terms that can arise in a
minimalistic model of continuum optomechanics. The simplest choice,
introduced in the beginning, would be identified as $\tilde{g}_{0}=g_{0}^{+++}$.
Even and odd terms cannot be present simultaneously, unless inversion
symmetry is broken. As remarked above, one can choose the definition
of the 1D field $\hat{u}(x)$ to select either the ``even'' or the
``odd'' representation. Note that the constants have different physical
dimensions (e.g. $g_{0}^{++-}$ is of dimensions $m^{3/2}{\rm Hz}$).

Interaction terms with derivatives would also arise by starting from
the microscopic theory, keeping the dispersion ($k$-dependence) of
the coupling, and translating from $k$-space into real space. In
general, this would yield derivatives of any order. Here, our aim
was to keep the leading terms. These are sufficient to retain a qualitatively
important feature: A model based on Eq.~(\ref{InteractionSimpleVersion})
would predict that the forward- and backward-scattering amplitudes
are equal (set by the single coupling constant in such a model). In
reality, that is not the case, and this fact is taken into account
properly by considering the derivatives.

Is our list complete? To answer this, let us discuss the ``even''
sector only, without loss of generality. In this sector, we went up
to second order in the derivatives, keeping terms such as $\left(\partial_{x}\hat{a}^{\dagger}\right)\hat{a}\left(\partial_{x}\hat{u}\right)$.
Why did we not consider second derivatives of individual fields, like
$\hat{a}^{\dagger}\hat{a}\left(\partial_{x}^{2}\hat{u}\right)$? The
answer is that these can indeed be present. However, a simple integration
by parts will transform those terms into a combination of the terms
that we already listed.

Beyond the interaction, the Hamiltonian contains the unperturbed energy
of the photons, $\hat{H}_{a}=\int dk\,\hbar\omega(k)\hat{a}^{\dagger}(k)\hat{a}(k)\,,$
and likewise $\hat{H}_{b}$ for the phonons with their dispersion
$\Omega(k)$. In real space, the same term could be written as $\hat{H}_{a}=\hbar\int dx\,\hat{a}^{\dagger}(x)\omega(-i\partial_{x})\hat{a}(x)\,,$
where $\omega(-i\partial_{x})$ applied to $e^{ikx}$ will reproduce
$\omega(k)$.

The resulting coupled continuum optomechanical Heisenberg\textcolor{blue}{{}
}equations of motion take the form:

\begin{eqnarray}
\partial_{t}\hat{a} & = & -i\omega(-i\partial_{x})\hat{a}+i\tilde{g}_{0}\hat{a}(\hat{b}+\hat{b}^{\dagger})\,\label{eq:eq_motion_a}\\
\partial_{t}\hat{b} & = & -i\Omega(-i\partial_{x})\hat{b}+i\tilde{g}_{0}\hat{a}^{\dagger}\hat{a}\,.\label{eq:eq_motion_b}
\end{eqnarray}
Here, eq. (4) and (5) are expressed with the simple interaction. More
generally, the interaction may be comprised of a linear combination
of terms in Table I. For example, the term $i\tilde{g}_{0}\hat{a}(\hat{b}+\hat{b}^{\dagger})$
in Eq. (\ref{eq:eq_motion_a}) becomes

\begin{eqnarray}
ig_{0}^{+++}\hat{a}\hat{u}-ig_{0}^{--+}\partial_{x}(\hat{u}\partial_{x}\hat{a})\nonumber \\
-ig_{0}^{-+-}\partial_{x}(\hat{a}\partial_{x}u)+ig_{0}^{-+-*}\left(\partial_{x}\hat{a}\right)\left(\partial_{x}\hat{u}\right)
\end{eqnarray}
when even couplings are considered. Likewise, term $i\tilde{g}_{0}\hat{a}^{\dagger}\hat{a}$
of Eq. (\ref{eq:eq_motion_b}) becomes

\begin{eqnarray}
ig_{0}^{+++}\hat{a}^{\dagger}\hat{a}+ig_{0}^{--+}(\partial_{x}\hat{a}^{\dagger})(\partial_{x}\hat{a})\nonumber \\
-ig_{0}^{-+-}\partial_{x}((\partial_{x}\hat{a}^{\dagger})\hat{a})-ig_{0}^{-+-*}\partial_{x}(\hat{a}^{\dagger}(\partial_{x}\hat{a}))
\end{eqnarray}
The real-space formulation developed here, with the complete list
of interactions derived above, will be especially powerful for considering
the effects of nonlinearities and of spatial inhomogeneities (whether
due to disorder or structure formation). No assumptions about the
fields peaking around a certain wavevector have been employed, nor
are we required to introduce a multitude of photon fields for cases
like forward scattering. The classical version of these nonlinear
equations can readily be solved by using split-step Fourier techniques.

\section{Dissipation and Quantum Noise}

To discuss the dissipation and the associated quantum and thermal
noise, we employ the well-known input-output formalism and adapt it
suitably to the continuum case. If we assume the photon loss rate
to be $\kappa$, then the equation of motion contains additional terms

\begin{equation}
\partial_{t}\hat{a}(x,t)=\ldots-\frac{\kappa}{2}\hat{a}(x,t)+\sqrt{\kappa}\hat{a}_{{\rm in}}(x,t)\,,
\end{equation}
where the vacuum noise field $\hat{a}_{{\rm in}}$ obeys the commutation
relation $\left[\!\right.\hat{a}_{{\rm in}}(x,t),\hat{a}_{{\rm in}}^{\dagger}(x',t')\left.\!\right]=\delta(x-x')\delta(t-t')$
and has the correlators $\left<\!\right.\hat{a}_{{\rm in}}(x,t)\hat{a}_{{\rm in}}^{\dagger}(x',t')\left.\!\right>=\delta(x-x')\delta(t-t')$
and $\left<\!\right.\hat{a}_{{\rm in}}^{\dagger}(x',t')\hat{a}_{{\rm in}}(x,t)\left.\!\right>=0$.
These ensure that the commutator of $\hat{a}$ is preserved, i.e.
the vacuum noise is constantly being replenished to offset the losses.

The mechanical field can be treated likewise, with a damping rate
$\Gamma$ in place of $\kappa$, and with the additional contribution
of thermal noise: $\left<\!\right.\hat{b}_{{\rm in}}(x,t)\hat{b}_{{\rm in}}^{\dagger}(x',t')\left.\!\right>=(\bar{n}_{{\rm th}}+1)\delta(x-x')\delta(t-t')$
and $\left<\!\right.\hat{b}_{{\rm in}}^{\dagger}(x',t')\hat{b}_{{\rm in}}(x,t)\left.\!\right>=\bar{n}_{{\rm th}}\delta(x-x')\delta(t-t')$.
Here $\bar{n}_{{\rm th}}=(\exp(\hbar\Omega/k_{B}T)-1)^{-1}$ is the
Bose occupation at temperature $T$. For simplicity, we assume that
this can be evaluated at some fixed frequency $\Omega$, since the
phonon dispersion $\Omega(k)$ is usually nearly flat in the most
important applications.

\section{Boundary conditions}

We have now collected all the ingredients for continuum optomechanics,
except the driving and the boundary conditions. A laser injecting
light of amplitude $\alpha_{{\rm in}}$ at point $x$ would be described
by an additional term $\sqrt{\kappa_{{\rm ex}}}\alpha_{{\rm in}}(x,t)$
in the equations of motion, with $\alpha_{{\rm in}}(x,t)=\alpha_{{\rm in}}(x)e^{-i\omega_{L}t}$
for a continuous wave excitation. Here $\kappa_{{\rm ex}}$ is the
coupling to the field mode that is populated by the laser photons,
and $\hbar\omega_{L}\left|\alpha_{{\rm in}}(x,t)\right|^{2}$ would
be the power per unit length impinging on the waveguide at position
x. This description is appropriate for illumination from the side,
which is feasible (and analogous to standard cavity optomechanis)
but atypical in experiments. 

More commonly, light is injected at the waveguide entrance. In that
case, we consider a half-infinite system, starting at $x=0$ and extending
to the right (Fig.~\ref{FigBasicSetup}d). The boundary at $x=0$
must be such that incoming waves (including the quantum vacuum noise)
are perfectly launched into the waveguide as right-going waves, while
left-moving waves exit without reflection. For the simplest case of
a constant photon velocity $c$, we need to prescribe the right-going
amplitude at $x=0$,

\begin{equation}
\partial_{t}\hat{a}(0,t)-c\partial_{x}\hat{a}(0,t)=-\sqrt{\frac{2}{c}}\partial_{t}(\alpha_{{\rm in}}(t)+\hat{a}_{{\rm in}}(t))\,,\label{eq:BoundaryCondition}
\end{equation}
where the ingoing quantum noise has the correlator

\begin{equation}
\left\langle \hat{a}_{{\rm in}}(t)\hat{a}_{{\rm in}}^{\dagger}(0)\right\rangle =\delta(t)
\end{equation}
while $\left\langle \hat{a}_{{\rm in}}^{\dagger}(t)\hat{a}_{{\rm in}}(0)\right\rangle $
vanishes. Eq.~(\ref{eq:BoundaryCondition}) is valid also in the
presence of dissipation. The solution of the free wave equation $\partial_{t}^{2}\hat{a}-c^{2}\partial_{x}^{2}\hat{a}=0$
with the boundary condition (\ref{eq:BoundaryCondition}) is

\begin{equation}
\hat{a}(x,t)=\hat{a}_{\rightarrow}(x-ct)+\hat{a}_{\leftarrow}(x+ct)
\end{equation}
where the right-moving field is set by $\hat{a}_{{\rm in}}(t)$:

\begin{equation}
\hat{a}_{\rightarrow}(x)=\hat{a}_{{\rm in}}(x/c)/\sqrt{2c}
\end{equation}
The left-moving field is an independent fluctuating field. The correlator
of the right-movers is

\begin{eqnarray*}
\left\langle \hat{a}_{\rightarrow}(x)\hat{a}_{\rightarrow}^{\dagger}(x')\right\rangle  & = & \frac{1}{2c}\left\langle \hat{a}_{{\rm in}}(x/c)\hat{a}_{{\rm in}}^{\dagger}(x'/c)\right\rangle \\
 & = & \frac{1}{2c}\delta(\frac{x-x'}{c})=\frac{1}{2}\delta(x-x')\\
\end{eqnarray*}
and the same result holds for the left-movers, such that the full
equal-time correlator of the $\hat{a}(x)$ field is set by $\delta(x-x')$.

\section{Rotating Frame and Linearized Description}

We can switch to a rotating frame, $\hat{a}^{{\rm old}}(x,t)=\hat{a}^{{\rm new}}(x,t)\cdot e^{i(k_{L}x-\omega_{L}t)}$,
where $\omega_{L}=\omega(k_{L})$ is the laser frequency. For brevity,
we drop the superscript 'new', i.e. all $\hat{a}$ are now understood
to be in the rotating frame. We then have, in the photon equation
of motion, after employing $\omega(-i\partial_{x})e^{ik_{L}x}=e^{ik_{L}x}\omega(k_{L}-i\partial_{x})$:

\begin{equation}
\partial_{t}\hat{a}=-i\tilde{\omega}(-i\partial_{x})\hat{a}+\ldots\,\label{eq:RotatingFrame}
\end{equation}
For brevity, we define $\tilde{\omega}(-i\partial_{x})=\omega(k_{L}-i\partial_{x})-\omega_{L}$.
If only modes close to $k_{L}$ are present, this may be expanded
using the group velocity $v=d\omega(k_{L})/dk$:

\begin{equation}
\partial_{t}\hat{a}=-v\partial_{x}\hat{a}+\ldots\,.\label{eq:LinearizedDispersion}
\end{equation}
The equation for the phonon field remains unaffected by this change.

We can linearize the equations in the standard way (see Supplementary
Material), setting $\beta(x)=\left\langle \hat{b}(x)\right\rangle $
and $\alpha(x)=\left\langle \hat{a}(x)\right\rangle $ for the steady-state
solution, and $\delta\hat{b}=\hat{b}-\beta$ and $\delta\hat{a}=\hat{a}-\alpha$
for the fluctuations. Then we obtain, for the simplest interaction
term:

\begin{eqnarray}
\partial_{t}\delta\hat{a} & = & -i\tilde{\omega}(-i\partial_{x})\delta\hat{a}+i\tilde{g}(x)(\delta\hat{b}+\delta\hat{b}^{\dagger})+i\tilde{g}_{\beta}(x)\delta\hat{a}+\ldots\label{eq:LinEqA}\\
\partial_{t}\delta\hat{b} & = & -i\Omega(-i\partial_{x})\delta\hat{b}+i(\tilde{g}(x)\delta\hat{a}(x)^{\dagger}+\tilde{g}^{*}(x)\delta\hat{a}(x))+\ldots\label{eq:LinEqB}
\end{eqnarray}
Here we introduced the linearized coupling $\tilde{g}(x)\equiv\tilde{g}_{0}\alpha(x)$,
as well as the shift $\tilde{g}_{\beta}(x)\equiv\tilde{g}_{0}(\beta(x)+\beta^{*}(x))$.
The omitted terms ($\ldots$) in Eqs. (\ref{eq:LinEqA}) and (\ref{eq:LinEqB})
contain the dissipation and fluctuations, in the same form as above
(only with $\hat{a}\mapsto\delta\hat{a}$, and the same for the phonons).
The boundary conditions for the fluctuations $\delta\hat{a}$ do not
contain any laser driving any more; i.e. we would have Eq.~(\ref{eq:BoundaryCondition})
for $\delta\hat{a}$, but without the laser amplitude $\alpha_{{\rm in}}$.

\section{Continuum limit for optomechanical arrays}

\begin{figure}
\includegraphics[width=1\columnwidth]{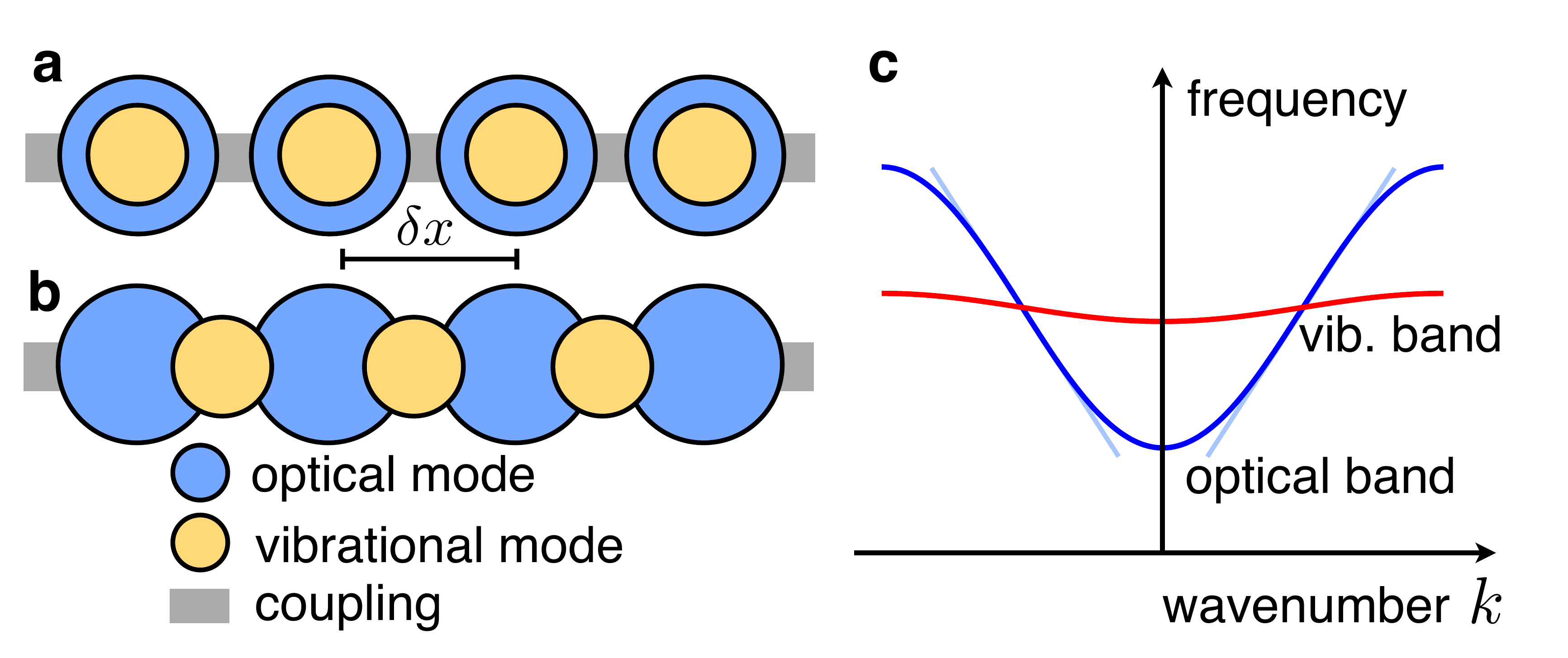}

\protect\caption{\label{OptoMechArray}(a) Schematic of a 1D optomechanical array,
with discrete localized optical and vibrational modes that are coupled
locally. (b) An alternative situation, where phonons couple to photon
tunneling between sites. (c) Bare bandstructure for phonons (red)
and photons (blue; shifted by an offset that is determined by the
pump laser frequency, using a rotating frame). The plot is shown for
zero optomechanical coupling.}

\end{figure}

In an optomechanical array, discrete localized optical and vibrational
modes are coupled to each other via the optomechanical interaction
$-\hbar g_{0}\hat{a}_{j}^{\dagger}\hat{a}_{j}(\hat{b}_{j}+\hat{b}_{j}^{\dagger})$,
see Fig.~\ref{OptoMechArray}. In addition, the photon and phonon
modes $\hat{a}_{j}$ and $\hat{b}_{j}$ are coupled by tunneling between
neighboring sites. For the photons, in a 1D array, this is described
by the tight-binding Hamiltonian $-\hbar\sum_{j,l}J_{l}\hat{a}_{j+l}^{\dagger}\hat{a}_{j}+{\rm h.c.}$.
Here $J_{l}$ is the tunnel coupling connecting any two sites $j$
and $j+l$. The resulting dispersion relation for the optical tight-binding
band is $\omega(k)=-\sum_{l}e^{-ikl\delta x}J_{l}$, where we already
introduced the lattice constant $\delta x$. For the phonons, an analogous
Hamiltonian holds, with a coupling constant $K_{l}$ and a resulting
phononic band $\Omega(k)$. 

The continuum theory will be a faithful approximation if only modes
of sufficiently long wavelengths (many lattice spacings) are excited.
The properly normalized way to identify localized modes with the continuum
fields is

\begin{equation}
\hat{a}_{j}=\hat{a}(j\delta x)\sqrt{\delta x},\,\hat{b}_{j}=\hat{b}(j\delta x)\sqrt{\delta x}\,.
\end{equation}
This ensures the validity of the commutator relations such as $\left[\hat{a}(x),\hat{a}^{\dagger}(x')\right]=\delta(x-x')$.
We then obtain

\begin{eqnarray}
\hat{H}_{{\rm int}}^{{\rm array}} & = & -\hbar g_{0}\sum_{j}\hat{a}_{j}^{\dagger}\hat{a}_{j}(\hat{b}_{j}+\hat{b}_{j}^{\dagger})\nonumber \\
 & \approx & \hat{H}_{{\rm int}}^{{\rm cont}}\,.
\end{eqnarray}
Here $\hat{H}_{{\rm int}}^{{\rm cont}}$ is the continuum version
of Eq.~(\ref{InteractionSimpleVersion}). For this simple local interaction,
none of the 'derivative-terms' appears. The present approximation
holds when the Hamiltonian acts on states where only long-wavelength
modes are excited. We can now relate the coupling constants for the
continuum and the discrete model:

\begin{equation}
\tilde{g}_{0}=g_{0}\sqrt{\delta x}\,.
\end{equation}
In taking the proper continuum limit, $\tilde{g}_{0}$ has to be kept
fixed, i.e. $g_{0}\sim1/\sqrt{\delta x}$ as $\delta x\rightarrow0$.
This is the expected physical behaviour, since $g_{0}\sim x_{{\rm ZPF}}$,
where $x_{{\rm ZPF}}=\sqrt{\hbar/(2m\Omega)}$ is the size of the
mechanical zero-point fluctuations of a discrete mechanical mode.
If this mode represents a piece of length $\delta x$ in a continuous
waveguide, its mass scales as $m=\delta x\cdot\rho$ {[}with $\rho$
the mass density{]}, such that $g_{0}$ grows in the manner discussed
above when $\delta x$ is sent to zero. Note that the continuum limit
also means keeping $\omega(k)$ and $\Omega(k)$ fixed in the relevant
wavelength range.

One can now also confirm that our treatment of quantum noise and dissipation
corresponds to the input-output formalism applied to the discrete
modes. For such modes, we would have $\dot{\hat{a}}_{j}=\ldots-\frac{\kappa}{2}\hat{a}_{j}+\sqrt{\kappa}\hat{a}_{j,{\rm in}}(t)$,
with $\left\langle \hat{a}_{j,{\rm in}}(t)\hat{a}_{j',{\rm in}}^{\dagger}(0)\right\rangle =\delta_{j,j'}\delta(t)$
and $\left[\hat{a}_{j,{\rm in}},\hat{a}_{j',{\rm in}}^{\dagger}\right]=\delta_{j,j'}$.
Setting $\hat{a}_{j,{\rm in}}(t)=\sqrt{\delta x}\hat{a}_{{\rm in}}(j\delta x,t)$,
this turns into the continuum expressions given above.

We turn back to the optomechanical interaction in the array. So far,
we had assumed a local interaction of the type $-\hbar g_{0}\hat{a}^{\dagger}\hat{a}(\hat{b}_{j}^{\dagger}+\hat{b}_{j})$.
However, it is equally possible to have an interaction that creates
phononic excitations during the photon tunneling process: $-\hbar g_{0}(\hat{a}_{j+1}^{\dagger}\hat{a}_{j}+{\rm h.c.})\hat{u}_{j}$,
where $\hat{u}_{j}\equiv\hat{b}_{j}+\hat{b}_{j}^{\dagger}$ describes
the displacement of a mode attached to the link between the sites
$j$ and $j+1$; see Fig.~\ref{OptoMechArray}b. It turns out that
such a coupling gives rise to 'derivative' terms in the continuum
model, see the Supplementary Material.

\section{Elementary processes for a single optical branch}

We briefly connect the real-space and $k$-space pictures to review
the elementary scattering processes. Translating the couplings in
table \ref{PossibleCouplings} to $k$-space, we arrive at the substitutions
$\hat{a}(x)\mapsto\hat{a}_{k}$, $\hat{a}^{\dagger}(x)\mapsto\hat{a}_{k+q}^{\dagger}$
and $\hat{u}(x)\mapsto\hat{u}_{q}$, with $\hat{u}_{q}=\hat{b}_{q}+\hat{b}_{-q}^{\dagger}$.
In addition, $\partial_{x}\hat{a}\mapsto ik\hat{a}_{k}$, $\partial_{x}\hat{a}^{\dagger}\mapsto-i(k+q)\hat{a}_{k+q}^{\dagger}$,
and $\partial_{x}\hat{u}\mapsto iq\hat{u}_{q}$. This yields the following
amplitude (for the example of the ``even'' sector) in front of the
resulting term $\hat{a}_{k+q}^{\dagger}\hat{a}_{k}\hat{u}_{q}$ in
the Hamiltonian:

\begin{equation}
-\hbar\left\{ g_{0}^{+++}+g_{0}^{--+}(k+q)k+g_{0}^{-+-}(k+q)q-g_{0}^{-+-*}kq\right\} 
\end{equation}
We can now specifically distinguish the amplitudes for forward-scattering
($q\approx0$):

\begin{equation}
g_{0F}=g_{0}^{+++}+g_{0}^{--+}k^{2}
\end{equation}
and backward-scattering ($q\approx-2k$):

\begin{equation}
g_{0B}=g_{0}^{+++}-k^{2}g_{0}^{--+}+2k^{2}(g_{0}^{-+-}+g_{0}^{-+-*})\,.
\end{equation}
Clearly it was important to keep more than the simplest interaction
term $g_{0}^{+++}$ in real-space to allow that these amplitudes are
different.

If only forward-scattering is considered, the situation is significantly
different from standard cavity optomechanics. The reason is that the
cavity allows us to introduce an asymmetry between Stokes and anti-Stokes
processes. This is absent here in forward-scattering, where phonons
of wavenumber $q$ can be emitted and absorbed equally likely, scattering
laser photons into a comb \cite{kang_tightly_2009,butsch_cw-pumped_2014,koehler_resolving_2016,wolff_cascaded_2016}
of sidebands $\omega_{L}\pm n\Omega$ with $\Omega=\Omega(q)$. Because
of this, basic phenomena in cavity optomechanics, like cooling or
state transfer, do not translate to the forward scattering case with
a single optical branch; these operation require asymmetry between
Stokes and anti-Stokes coupling processes. Dispersive symmetry breaking
is seldom accomplished in this geometry, as typical propagation lengths
are not adequate to resolve the wavevector difference between Stokes
and anti-Stokes phonon modes. 

In backward scattering, the situation is different, since either phonons
of wavenumber $q\cong2k_{L}$ are emitted (Stokes) or those of wavenumber
$q\cong-2k_{L}$ are absorbed (anti-Stokes). This can result in cooling
of $-2k_{L}$ phonons and amplification of $+2k_{L}$ phonons. The
latter process amounts to stimulated backward Brillouin scattering,
amplifying any counterpropagating beam.

\section{Multiple optical branches}

The useful Stokes/anti-Stokes asymmetry can be re-introduced into
forward scattering by considering multiple optical branches. These
might be different transverse optical modes. In that case, the (simplest)
interaction is

\begin{equation}
-\hbar\sum_{j,l}\int dx\,\tilde{g}_{0}(j,l)\hat{a}_{j}^{\dagger}(x)\hat{a}_{l}(x)(\hat{b}(x)+\hat{b}^{\dagger}(x))\,.
\end{equation}
Here $\tilde{g}_{0}(j,l)$ describes the bare coupling for scattering
from branch $l$ to $j$, with $\tilde{g}_{0}^{*}(l,j)=\tilde{g}_{0}(j,l)$,
and $[\hat{a}_{j}(x),\hat{a}_{l}^{\dagger}(x')]=\delta_{jl}\delta(x-x')$.
Analogous expressions can be written down for the other interactions
of table \ref{PossibleCouplings}.

For the case of two branches, there will be forward-scattering of
photons $k_{L}\mapsto k_{L}+q$ between the branches, by either absorbing
a phonon of wavenumber $q$ or emitting one of wavenumber $-q$. In
the linearized Hamiltonian, the inter-branch scattering process is
described by

\begin{equation}
-\hbar\int dx\,(\tilde{g}_{21}(x)\delta\hat{a}_{2}^{\dagger}(x)+{\rm h.c.})(\delta\hat{b}(x)+\delta\hat{b}^{\dagger}(x))\,,\label{TwoBranchesLinearized}
\end{equation}
with $\tilde{g}_{21}(x)=\tilde{g}_{21}e^{ik_{L}x}$, where $\tilde{g}_{21}=\tilde{g}_{0}(2,1)\alpha_{1}$.
In momentum space, this turns into

\begin{equation}
-\hbar\int dq\,\tilde{g}_{21}\delta\hat{a}_{2}[k_{L}+q]^{\dagger}(\delta\hat{b}[q]+\delta\hat{b}[-q]^{\dagger})+{\rm h.c.}
\end{equation}

\section{Interband scattering: Weak coupling}

We first treat the weak coupling limit for scattering between different
optical bands, which has been discussed widely in the literature and
is known under various names such as stimulated Brillouin scattering
(SBS) or stimulated Raman-like scattering (see Suppl. Material for
a discussion of naming conventions). It is a widely studied regime
of continuum optomechanical coupling, with a long history in the context
of nonlinear optics \cite{chiao_stimulated_1964,shen_theory_1965,boyd_nonlinear_2013,agrawal_nonlinear_2012}.
The phonon fields are assumed to have far shorter decay lengths than
the optical waves, which is frequently satisfied by experimental systems.
In this limit, the nonlinear optical susceptibility induced by optomechanics
can be approximated as local, greatly simplifying the spatio-temporal
dynamics. For clarity, we term this regime the 'Brillouin-limit'. 

\textit{\emph{To connect our continuum optomechanical framework with
Brillouin or Raman interactions, we start from Eq.~(\ref{TwoBranchesLinearized})}}\emph{,}
for two optical branches\emph{.} Just as in Eqs. (\ref{eq:RotatingFrame})
and (\ref{eq:LinearizedDispersion}), we introduce rotating frames
and linearize the dispersion relations. Then, we obtain: 
\begin{eqnarray}
\partial_{x}\left\langle \delta\hat{a}_{2}\right\rangle  & = & i(\tilde{g}_{12}/v_{2})\left\langle \delta\hat{b}^{\dagger}\right\rangle -(\gamma_{2}/2)\left\langle \delta\hat{a}_{2}\right\rangle ,\label{eq:SpatialEqMotionFields_a2}\\
\partial_{x}\left\langle \delta\hat{b}\right\rangle  & = & i(\tilde{g}_{12}/v_{b})\left\langle \delta\hat{a}_{2}^{\dagger}\right\rangle -(\gamma_{b}/2)\left\langle \delta\hat{b}\right\rangle .\label{eq:SpatialEqMotionFields_b}
\end{eqnarray}
Here, $\gamma_{2}\equiv\kappa_{2}/v_{2}$ and $\gamma_{b}\equiv\Gamma/v_{b}$
represent the spatial power decay rate of the photon (phonon) fields,
i.e. the inverse decay length. Since the spatial decay rate of sound
($\gamma_{b}$) is typically much larger than that of light ($\gamma_{2}$),
the phonon field is generated locally: $\partial_{x}\left\langle \delta\hat{b}\right\rangle \approx0$.
This allows to express the mechanical amplitude in terms of the light
field, which yields: 
\begin{eqnarray}
\partial_{x}\left\langle \delta\hat{a}_{2}\right\rangle  & = & \frac{|\tilde{g}_{12}|^{2}}{v_{1}\Gamma}\left\langle \delta\hat{a}_{2}\right\rangle -(\gamma_{2}/2)\left\langle \delta\hat{a}_{2}\right\rangle ,
\end{eqnarray}
We can now cast this result in terms of traveling-wave optical powers
$P_{1}$ and $P_{2}$, with $P_{1}=\hbar\omega_{1}v_{1}\left|\alpha_{1}\right|^{2}$,
$P_{2}\cong\hbar\omega_{2}v_{2}\left|\left\langle \delta\hat{a}_{2}\right\rangle \right|^{2}$,
and $P_{b}\cong\hbar\Omega v_{b}\left|\left\langle \delta\hat{b}\right\rangle \right|^{2}$.
Here we assumed the small signal limit, i.e. $\alpha_{2}=0$, $\beta=0$,
and $\alpha_{1}$ is large. We see that $P_{2}$ is exponentially
amplified according to 
\begin{eqnarray}
\frac{\partial P_{2}}{\partial x} & = & G_{B}P_{1}P_{2}-\gamma_{2}P_{2},
\end{eqnarray}
where $G_{B}\equiv4|\tilde{g}_{o}({1,2})|^{2}/(v_{1}v_{2}\Gamma\hbar\omega_{1})$
is the Brillouin gain coefficient \cite{agrawal_nonlinear_2012,rakich_giant_2012}.
For alternative derivations in the context of nonlinear optics and
Brillouin photonics, see Refs.~\cite{agrawal_nonlinear_2012,boyd_nonlinear_2013};
for discussion of the induced nonlinear optical susceptibility see
Suppl. Material. This relationship between $G_{B}$ and $\tilde{g}_{o}({1,2})$
permits us to leverage established methods for calculation of the
optomechanical coupling in both translationally invariant \cite{rakich_giant_2012,qiu_stimulated_2013,wolff_stimulated_2015,sipe_hamiltonian_2016}
and periodic \cite{qiu_stimulated_2012} nano-optomechanical systems.
In the Brilloin limit, a range of complex spatio-temporal phenomena
have been studied \cite{agrawal_nonlinear_2012,boyd_nonlinear_2013}\emph{.}\textit{}

\section{Strong coupling in the 'Coherent-Phonon Limit'}

The case opposite to the 'Brillouin limit', that we just discussed,
is the situation of a large phonon coherence length. This might be
termed the 'coherent phonon limit'. In this much less explored limit,
a large variety of interesting classical and quantum phenomena can
be expected to appear, as the system acquires a much higher degree
of coherence and nonlocality. Quantum states can then be swapped between
the light field and the phonon field, which can lead to applications
like opto-acoustic data storage in a fibre \cite{zhu_stored_2007}.
We now consider the situation where creation of a photon in the second
branch is accompanied by \emph{absorption} of a phonon, instead of
the emission that would lead to amplification. This leads to a modified
version of Eqs.~(\ref{eq:SpatialEqMotionFields_a2}) and (\ref{eq:SpatialEqMotionFields_b}):

\begin{eqnarray}
\partial_{x}\left\langle \delta\hat{a}_{2}\right\rangle  & = & i(\tilde{g}_{12}/v_{2})\left\langle \delta\hat{b}\right\rangle -(\gamma_{2}/2)\left\langle \delta\hat{a}_{2}\right\rangle ,\label{eq:SpatialEqMotionFields_a2_swap}\\
\partial_{x}\left\langle \delta\hat{b}\right\rangle  & = & i(\tilde{g}_{12}^{*}/v_{b})\left\langle \delta\hat{a}_{2}\right\rangle -(\gamma_{b}/2)\left\langle \delta\hat{b}\right\rangle .\label{eq:SpatialEqMotionFields_b_swap}
\end{eqnarray}
That can be recast as a matrix equation

\begin{equation}
\partial_{x}\phi=M\phi\,,
\end{equation}
where the vector $\phi$ contains the fields, $\phi=\left(\left\langle \delta\hat{a}_{2}\right\rangle ,\left\langle \delta\hat{b}\right\rangle \right){}^{T}$,
and

\begin{equation}
M=\left(\begin{array}{cc}
-\gamma_{2}/2 & i\tilde{g}_{12}/v_{2}\\
i\tilde{g}_{12}^{*}/v_{b} & -\gamma_{b}/2
\end{array}\right)\,.
\end{equation}
This is a non-Hermitian matrix that can be diagonalized to obtain
the spatial evolution $\phi\sim e^{\lambda x}$. We find the eigenvalues

\begin{equation}
\lambda_{\pm}=\frac{1}{2}\left[-\bar{\gamma}\pm\sqrt{D}\right]\,,
\end{equation}
where $\bar{\gamma}=(\gamma_{2}+\gamma_{b})/2$ is the average spatial
decay rate, and $D=[(\gamma_{2}-\gamma_{b})/2]^{2}-4\left|\tilde{g}_{12}\right|^{2}/v_{2}v_{b}$.
A distinct oscillatory regime is reached when $D<0$, i.e.

\begin{equation}
\left|\tilde{g}_{12}\right|>\sqrt{v_{2}v_{b}}\frac{\left|\gamma_{2}-\gamma_{b}\right|}{4}\,.\label{eq:CoherentPhononThreshold}
\end{equation}
In that case, the eigenvalues attain an imaginary part, and the spatial
evolution becomes oscillatory. Interestingly, this sharp threshold
only depends on the \emph{difference} of spatial decay rates. In principle,
therefore, in an unconventional system where $\gamma_{2}$ and $\gamma_{b}$
are of the same order, this condition is much easier to fulfill than
when having to compare $\left|\tilde{g}_{12}\right|$ against the
total decay rate. Nevertheless, in order for the oscillations to be
observed in practice, in addition the decay length should be larger
than the period of oscillations. This will be true when $\left|{\rm Re}\lambda\right|\ll\left|{\rm Im}\lambda\right|$,
which can be approximated as 
\begin{equation}
\left|\tilde{g}_{12}\right|\gg\sqrt{v_{2}v_{b}}\bar{\gamma}/2\,.\label{eq:StrongCoupling}
\end{equation}
We will term this the \emph{``strong coupling regime}'' for continuum
optomechanics. It is in spirit similar to the strong coupling regime
of cavity optomechanics \cite{aspelmeyer_cavity_2014}, although the
dependence on the velocities introduces a new element. If this more
demanding condition (\ref{eq:StrongCoupling}) is fulfilled, then
the coupling is also automatically larger than the threshold (\ref{eq:CoherentPhononThreshold})
given above. 

To interpret this condition, note that usually $\bar{\gamma}$ is
dominated by the phonon decay $\gamma_{b}=\Gamma/v_{b}$. In that
case, we could also write $\left|\tilde{g}_{12}\right|\gg\sqrt{v_{2}/v_{b}}\Gamma/4$.
This shows that, at a fixed phonon decay rate $\Gamma$, smaller phonon
velocities make the strong coupling regime harder to reach.

\section{Experimental Overview}

\begin{figure}
\includegraphics[width=1\columnwidth]{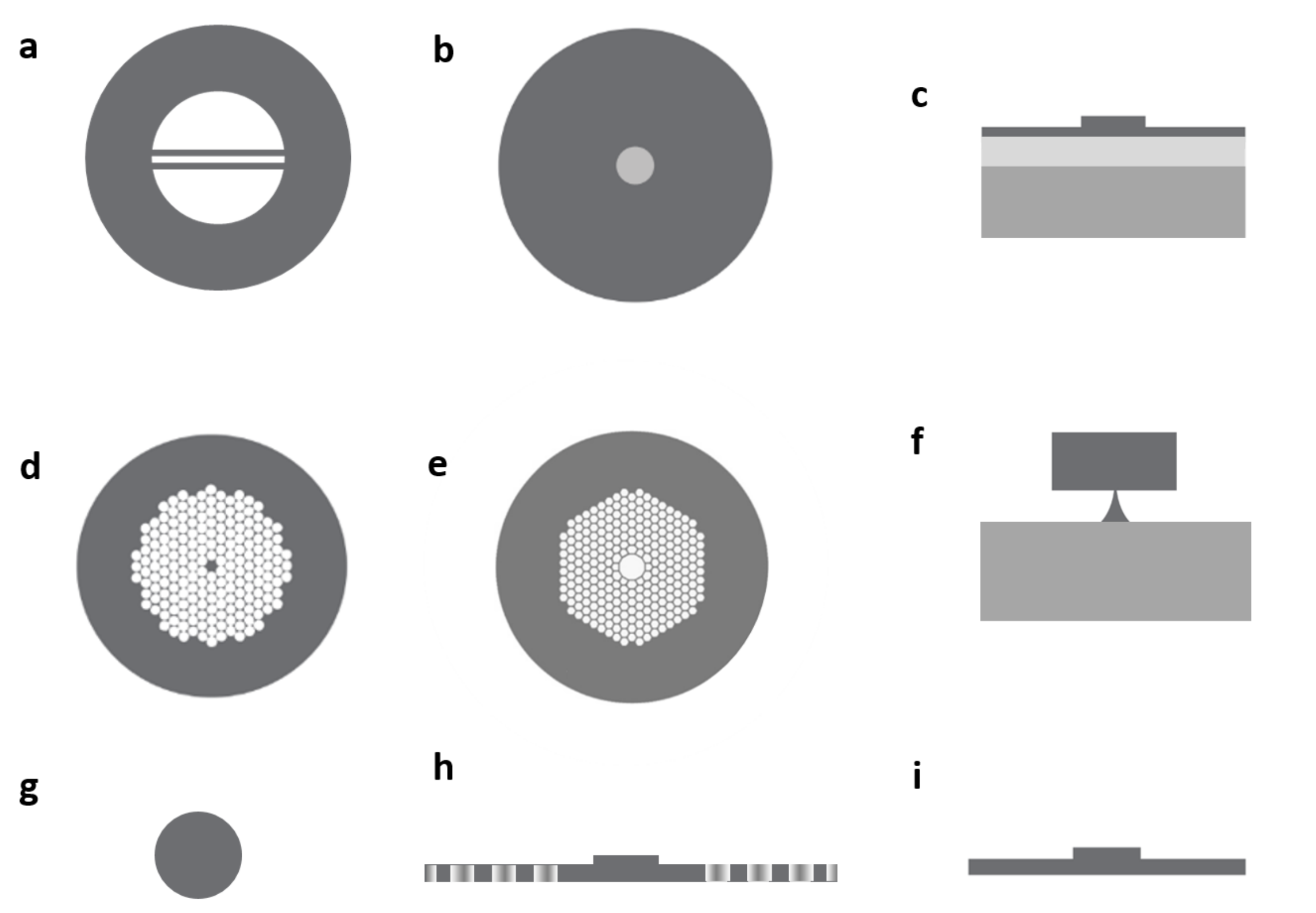}

\protect\protect\caption{\label{ExpSystems Overview} Overview of experimental systems including:
(a) Nanoweb fibre \cite{butsch_optomechanical_2012}; (b) Step-index
fibre \cite{abedin_observation_2005,behunin_long-lived_2015}; (c)
Ridge waveguide \cite{pant_-chip_2010}; (d) Crystal fibre \cite{kang_tightly_2009,beugnot_guided_2007,zhong_depolarized_2015};
(e) Hollow core photonic crystal fiber \cite{zhong_depolarized_2015,renninger_guided-wave_2016,renninger_nonlinear_2016,renninger_forward_2016-1};
(f) Nanowire silicon waveguide \cite{van_laer_interaction_2015,laer_net_2015};
(g) Silica nanowire fiber \cite{beugnot_brillouin_2014}; (h) Membrane
suspended phononic crystal waveguide \cite{shin_control_2015}; (e)
Membrane suspended silicon waveguide \cite{shin_tailorable_2013,kittlaus_large_2016}.}
\end{figure}

\begin{figure}
\includegraphics[width=1\columnwidth]{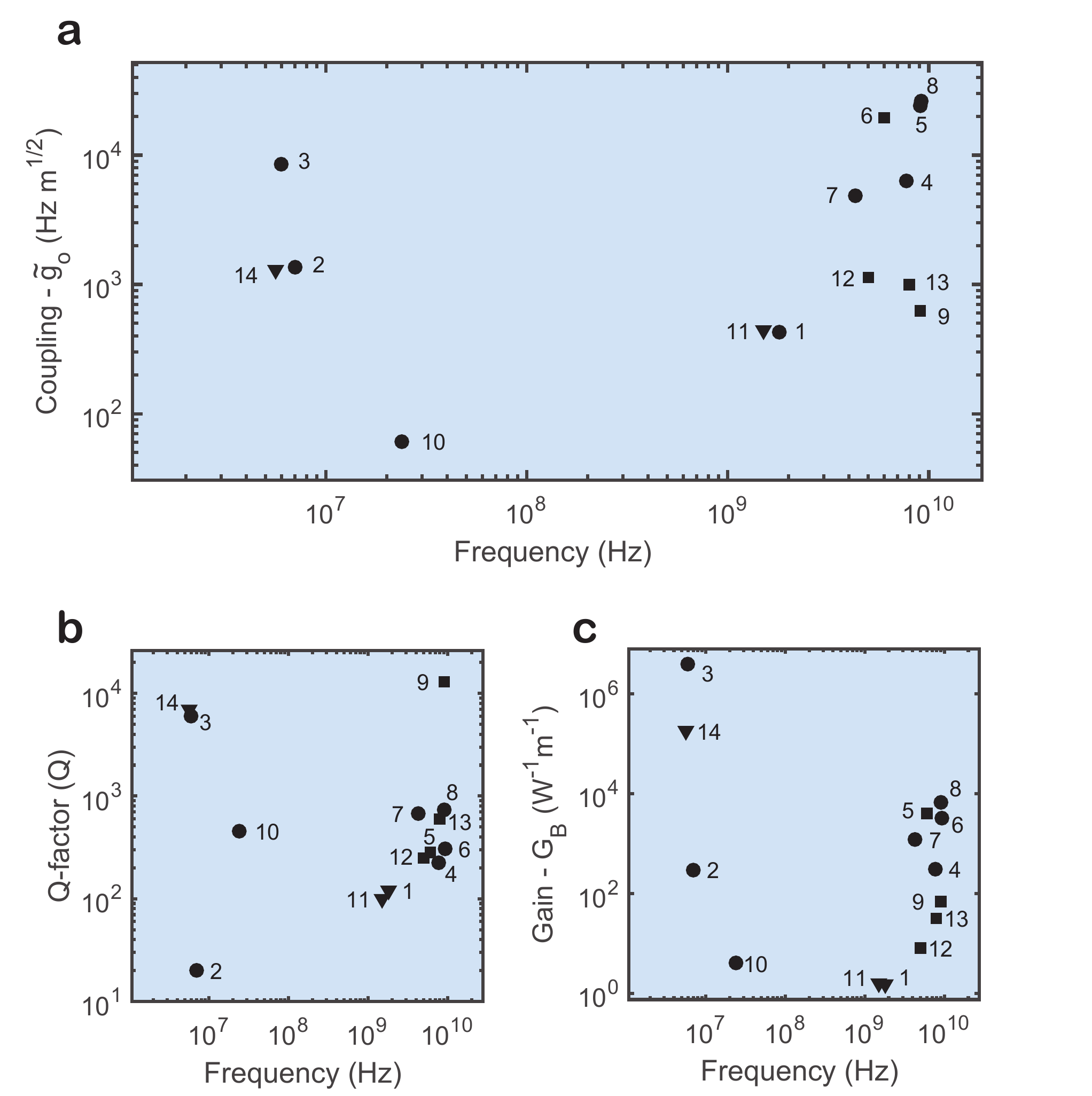}

\protect\protect\caption{\label{ScatterPlots} Analysis of experimental parameters for some
representative experimental systems. (a), (b) and (c) show estimated
continuum-optomechanical coupling constant, effective mechanical Q
factor, and Brillouin gain. The horizontal axis displays the phonon
frequency in each case. 1: Photonic crystal fibre (Kang et al 2009
\cite{kang_tightly_2009}); 2,3: ``Nanoweb'' fibre (2 Butsch et
al. 2012 \cite{butsch_optomechanical_2012}, 3 Butsch et al. 2014
\cite{butsch_cw-pumped_2014}); 4: Chalcogenide ridge waveguide (Pant
2010 \cite{pant_-chip_2010}); 5,7: Membrane suspended silicon waveguide
(5 Shin et al. 2013 \cite{shin_tailorable_2013}, 7 Kittlaus et al.
2016 \cite{kittlaus_large_2016}); 6,8: Silicon photonic nanowire
(6 Laer et al. 2015 \cite{van_laer_interaction_2015}, 8 Laer et al.
2015 \cite{laer_net_2015}); 9: Single-mode fibre (Behunin et al.
2015 \cite{behunin_long-lived_2015}); 10: Helium filled hollow-core
photonic crystal fiber (Renninger et al. 2016 \cite{renninger_nonlinear_2016});
11: Photonic crystal fibre (Kang et al. 2010 \cite{kang_all-optical_2010})
12: Silica nanowire fibre (Beugnot et al. 2014 \cite{beugnot_brillouin_2014});
13 Chalcogenide fibre (Abedin 2014 \cite{abedin_observation_2005}.);
14 ``Nanoweb'' fibre (Koehler et al. al. 2012 \cite{koehler_resolving_2016})}
\end{figure}

Coupling between continuous optical and phonon fields has been realized
in the context of nonlinear optics studies of Brillouin interactions.
These experimental systems, depicted in Fig.~\ref{ExpSystems Overview},
include step-index and micro-structured optical fibers \cite{abedin_observation_2005,beugnot_brillouin_2014,kang_tightly_2009,kang_all-optical_2010,butsch_optomechanical_2012,behunin_long-lived_2015,beugnot_guided_2007},
gas- and superfluid-filled photonic bandgap fibers\cite{zhong_depolarized_2015,renninger_guided-wave_2016,renninger_nonlinear_2016,renninger_forward_2016-1},
as well as chip-scale integrated optomechanical waveguide systems
\cite{pant_-chip_2010,shin_tailorable_2013,shin_control_2015,van_laer_interaction_2015,laer_net_2015,kittlaus_large_2016}.
To date, these studies have overwhelmingly focused on the Brillouin
related nonlinear optical phenomena \cite{abedin_observation_2005,beugnot_brillouin_2014,kang_tightly_2009,kang_all-optical_2010,butsch_optomechanical_2012,behunin_long-lived_2015,beugnot_guided_2007,renninger_guided-wave_2016,renninger_nonlinear_2016,renninger_forward_2016-1,pant_-chip_2010,shin_tailorable_2013,shin_control_2015,van_laer_interaction_2015,laer_net_2015,kittlaus_large_2016},
as well as noise processes \cite{shelby_guided_1985,zhong_depolarized_2015,elser_reduction_2006}.
However, it is also interesting to examine these systems through the
lens of continuum optomechanics. Figure \ref{ScatterPlots}a shows
the estimated continuum-optomechanical coupling strengths, extracted
using the Brillouin gain $G_{B}$, as derived in the previous section.
We see that couplings of between $10^{2}-10^{4}~\textup{Hz}\cdot\textup{m}^{1/2}$
have been realized using radiation pressure and (or) photo-elastic
coupling. These couplings are mediated by phonons with frequencies
between 10 MHz and 18 GHz depending on the type of interaction (intra-band
or inter-band) and the elastic wave that mediates the coupling.

\begin{figure}
\includegraphics[width=1\columnwidth]{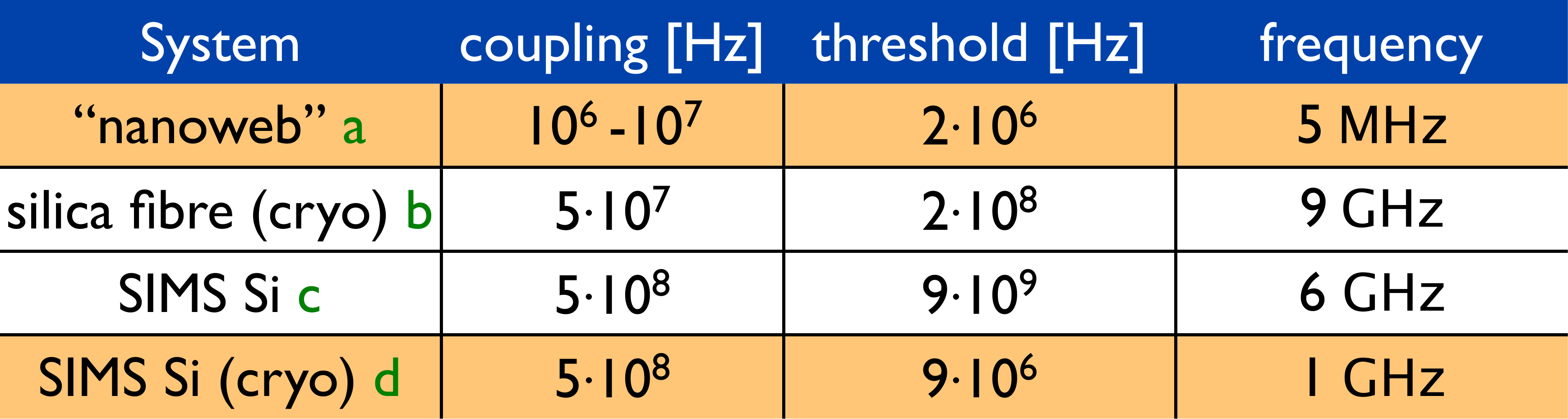}

\protect\caption{\label{fig:Possible-experimental-access}Possible experimental access
to the strong coupling regime: The coupling $\left|\tilde{g}_{12}\right|$
needs to be much larger than the 'threshold' $\sqrt{v_{2}/v_{b}}\Gamma/4$.
Estimated values for a: \cite{koehler_resolving_2016}, b: \cite{behunin_long-lived_2015},
c: stimulated intermodal scattering in silicon (room temp.), d: same
at $1K$, assuming an increase of $Q$ by a factor of 100.}

\end{figure}

The strength of the nonlinear optical susceptibility increases linearly
with phononic Q-factor. This is seen by comparing the effective phononic
Q-factors, plotted in Fig.~\ref{ScatterPlots}b with the peak Brillouin
gain of Fig.~\ref{ScatterPlots}c. We define the effective Q-factor
as the ratio of the mechanical frequency and the line-width. The effective
Q-factor is always smaller than the intrinsic phonon Q-factor due
to inhomogeneous broadening from variations in waveguide dimension
along the waveguide length \cite{wolff_brillouin_2016}.

A variety of single-band (intra-modal) and multi-band (inter-modal)
interactions have been demonstrated. These single-band processes include
intra-modal forward-SBS processes (also termed stimulated Raman-like
scattering) and backward-SBS processes; each process is denoted with
circular and square markers, respectively, in Fig.~\ref{ScatterPlots}.
Multi-band processes, generically termed inter-modal Brillouin processes,
are denoted by triangular markers in Fig. \ref{ScatterPlots}; their
classification and nomenclature is discussed in the Suppl. Material.

As discussed in the previous section, the phonon coherence length
has a significant impact on the spatio-temporal dynamics. Thus it
is important to note that, depending on the intrinsic Q-factor and
the type of phonon mode, the coherence length of the phonon can vary
dramatically. For instance, since intra-modal coupling is mediated
by phonons with vanishing group velocities ($\sim1m/s$) \cite{shin_tailorable_2013},
phonon coherence lengths are often less than 100 nm. Conversely, in
the cases of backward- or inter-modal (inter-band) coupling, the phonon
group velocities can approach the intrinsic sound velocity in the
waveguide material (e.g., $10^{4}m/s$). These higher velocity phonon
modes correspond to 10-50 micron coherence lengths at room temperatures,
but can be extended to milimeter length-scales at cryogenic temperatures\cite{behunin_long-lived_2015}.

Numerous nano-optomechanical devices have been proposed that have
the potential to yield increased coupling strengths \cite{qiu_stimulated_2012,van_laer_analysis_2014,sarabalis_guided_2016}.
Fig.~\ref{fig:Possible-experimental-access} indicates the prospects
for exploring the strong coupling regime discussed before.

\section{Conclusions}

We have established a connection between the continuum limit of optomechanical
arrays and Brillouin physics. Especially studies of (classical and
quantum) nonlinear dynamics will profit from our approach, where we
categorized the simplest coupling terms and derived the quantum Langevin
equations, including the noise terms and the correct boundary conditions.
Applications such as wavelength conversion, phonon-induced coherent
photon interactions and extensions to two-dimensional situations \cite{ruiz-rivas_dissipative_2016,butsch_optomechanical_2012-1}
can now be analyzed on the basis of this framework. As an example,
we have identified the strong coupling regime in continuum-optomechanical
systems and prospects for reaching it in the context of state-of-the-art
experimental systems.

\subsection*{Acknowledgements}

We thank Philip Russell and Andrey Sukhorukov for initial discussions
that helped inspire this project and for useful feedback on the manuscript.
We acknowledge support by an ERC Starting Grant (FM). P.T.R. acknowledges
support from the Packard Fellowship for Science and Engineering.

\section{Supplementary Material}

\subsection{Linearized Interaction}

We briefly review the (straightforward) route from the fully nonlinear
interaction to the linearized version, i.e. a quadratic Hamiltonian.
Assume a steady state solution has been found, with $\beta(x)=\left\langle \hat{b}(x)\right\rangle $
and $\alpha(x)=\left\langle \hat{a}(x)\right\rangle $. As is known
for standard cavity optomechanics, there might be more than one steady-state
solution, and formally there could be an infinity of solutions for
the continuum case. We have not explored this possibility further.

The deviations from this solution will now be denoted $\delta\hat{b}=\hat{b}-\beta$
and $\delta\hat{a}=\hat{a}-\alpha$. These are still fields. In contrast
to the standard single-mode case, we will keep the possibility that
$\beta(x)$ depends on position. 

On the Hamiltonian level, we now obtain a new 'linearized' (i.e. quadratic)
interaction term:

\begin{equation}
-\hbar\int dx\,\left[\tilde{g}(x)\delta\hat{a}^{\dagger}(x)+\tilde{g}^{*}(x)\delta\hat{a}(x)\right]\left[\delta\hat{b}(x)+\delta\hat{b}^{\dagger}(x)\right]\label{LinearizedHam}
\end{equation}
as well as a term

\begin{equation}
-\hbar\int dx\,\tilde{g}_{\beta}(x)\delta\hat{a}^{\dagger}(x)\delta\hat{a}(x)\,,
\end{equation}
which is a (possibly position-dependent) shift of the optical frequency.
Its counterpart in the cavity optomechanics case is often dropped
by an effective redefinition of the laser detuning.

In writing down Eq.~(\ref{LinearizedHam}), we have defined

\begin{eqnarray}
\tilde{g}(x) & \equiv & \tilde{g}_{0}\alpha(x)\\
\tilde{g}_{\beta}(x) & \equiv & \tilde{g}_{0}(\beta(x)+\beta^{*}(x))
\end{eqnarray}
The photon-enhanced continuum coupling strength $\tilde{g}(x)$ is
the direct analogue of the enhanced coupling $g=g_{0}\alpha$ in the
standard linearized cavity-optomechanical case. In contrast to $\tilde{g}_{0}$,
$\tilde{g}$ has the dimensions of a frequency. Likewise, $\tilde{g}_{\beta}$
is the static mechanical displacement, expressed as a resulting optical
frequency shift.

At this point, we have only started from the simplest kind of interaction,
Eq.~(\ref{InteractionSimpleVersion}), to obtain Eq.~(\ref{LinearizedHam}).
We will comment on the other terms of table \ref{PossibleCouplings}
below.

\subsection{Optomechanical Arrays: Derivative Terms in the continuum version
of the interaction}

In an optomechanical array, it is possible to have an interaction
that creates phononic excitations during the photon tunneling process:
$-\hbar g_{0}(\hat{a}_{j+1}^{\dagger}\hat{a}_{j}+{\rm h.c.})\hat{u}_{j}$,
where $\hat{u}_{j}\equiv\hat{b}_{j}+\hat{b}_{j}^{\dagger}$ describes
the phonon displacement of a mode attached to the link between the
sites $j$ and $j+1$. Here we describe how this can give rise to
the canonical derivative terms when switching to a continuum description.

Switching from the discrete lattice model to the continuum model,
we replace

\begin{equation}
\hat{a}_{j+1}^{\dagger}\hat{a}_{j}\hat{u}_{j}\mapsto\delta x^{3/2}\hat{a}^{\dagger}(x+\delta x/2)\hat{a}(x-\delta x/2)\hat{u}(x)\,,
\end{equation}
where we chose coordinates so as to indicate that the phonon mode
$\hat{u}$ is located halfway between the photon modes at $x\pm\delta x/2$.
A Taylor expansion of 

\begin{equation}
\hat{a}^{\dagger}(x+\delta x/2)\hat{a}(x-\delta x/2)+{\rm h.c.}
\end{equation}
yields

\begin{equation}
2\hat{a}^{\dagger}\hat{a}+\left(\frac{\delta x}{2}\right)^{2}\left\{ \left(\partial_{x}^{2}\hat{a}^{\dagger}\right)\hat{a}+\hat{a}^{\dagger}\left(\partial_{x}^{2}\hat{a}\right)-2\left(\partial_{x}\hat{a}^{\dagger}\right)\left(\partial_{x}\hat{a}\right)\right\} \,,\label{TaylorExpandedInteraction}
\end{equation}
where all fields are taken at position $x$. Two things are worth
noting here: First, all the first-order derivatives have disappeared
(they would have violated inversion symmetry!). Second, we have obtained
second-order derivatives of the photon field. If we want to turn this
into our ``canonical'' choice of coupling terms (table \ref{PossibleCouplings}),
we have to integrate by parts, in which case derivatives may also
act on $\hat{u}(x)$. This turns $\left\{ \left(\partial_{x}^{2}\hat{a}^{\dagger}\right)\hat{a}+\hat{a}^{\dagger}\left(\partial_{x}^{2}\hat{a}\right)\right\} \hat{u}$
into:

\begin{equation}
-2(\partial_{x}\hat{a}^{\dagger})(\partial_{x}\hat{a})\hat{u}-\left[\left(\partial_{x}\hat{a}^{\dagger}\right)\hat{a}+\hat{a}^{\dagger}\left(\partial_{x}\hat{a}\right)\right]\left(\partial_{x}\hat{u}\right)\,.
\end{equation}
Combining this with the other terms resulting from Eq.~(\ref{TaylorExpandedInteraction}),
one arrives at the interaction expressed completely in the canonical
way.

\subsection{Nonlinear susceptibility}

We briefly discuss how, starting from the linearized Eq.~(\ref{eq:SpatialEqMotionFields_b}),
we can obtain the effective third-order nonlinear photon susceptibility
induced by the interaction with the phonons. We slightly generalize
this equation, by adding a possible detuning between the mechanical
frequency $\Omega$ and the transition frequency $\Omega_{o}$ between
the two optical branches:

\begin{equation}
\partial_{x}\left\langle \delta\hat{b}\right\rangle =i[(\Omega-\Omega_{o})/v_{b}]\left\langle \delta\hat{b}\right\rangle +i(\tilde{g}_{12}/v_{b})\left\langle \delta\hat{a}_{2}^{\dagger}\right\rangle -(\gamma_{b}/2)\left\langle \delta\hat{b}\right\rangle .
\end{equation}
Solving for the steady state $\partial_{x}\left\langle \delta\hat{b}\right\rangle $
and inserting into the photon equation of motion, Eq.~(\ref{eq:SpatialEqMotionFields_a2}),
we obtain:

\begin{equation}
\partial_{x}\left\langle \delta\hat{a}_{2}\right\rangle =-\textcolor{red}{{\normalcolor \frac{i}{v_{2}}}}\frac{\left|\tilde{g}_{0}(1,2)\right|^{2}\left|\alpha_{1}\right|^{2}\textcolor{red}{{\normalcolor \left\langle \delta\hat{a}_{2}\right\rangle }}}{\Omega-\Omega_{0}-i\frac{\Gamma}{2}}-\frac{\gamma_{2}}{2}\left\langle \delta\hat{a}_{2}\right\rangle \,.
\end{equation}

We can express this as

\begin{equation}
\partial_{x}\left\langle \delta\hat{a}_{2}\right\rangle =i\gamma_{{\rm nonlin}}^{(3)}\left|\alpha_{1}\right|^{2}\textcolor{red}{{\normalcolor \left\langle \delta\hat{a}_{2}\right\rangle }}-\frac{\gamma_{2}}{2}\left\langle \delta\hat{a}_{2}\right\rangle \,,
\end{equation}
with the effective nonlinear susceptibility

\begin{equation}
\gamma_{{\rm nonlin}}^{(3)}\textcolor{red}{{\normalcolor (\Omega)}}=-\textcolor{red}{{\normalcolor \frac{1}{v_{2}}}}\frac{\left|\tilde{g}_{0}(1,2)\right|^{2}}{\Omega-\Omega_{0}-i\frac{\Gamma}{2}}\,.
\end{equation}
Using $P_{1}=\hbar\omega_{1}v_{1}\left|\alpha_{1}\right|^{2}$ and
$P_{2}\cong\hbar\omega_{2}v_{2}\left|\left\langle \delta\hat{a}_{2}\right\rangle \right|^{2}$
to cast Eq. 47 in the form of Eq. 28, one finds that the frequency
dependent gain is related to the nonlinear susceptibility as $G_{{\rm B}}(\Omega)=-2\cdot{\rm Im}\{\gamma_{{\rm nonlin}}^{(3)}(\Omega)\}(\hbar\omega_{1}v_{1})^{-1}$.

\subsection{Types of Brillouin interactions}

Here, we elucidate some naming conventions used in the Brillouin literature,
and we explain how these names relate to the classifications that
we use in this paper. These include (i) forward intra-band scattering
processes, where incident and scattered light-fields co-propagate
in the same optical mode, (ii) backward intra-band scattering processes,
where the incident and scattered light-fields counter-propagate, as
well as (iii) inter-band scattering processes, which generically describe
processes that involve coupling between guided optical modes with
distinct dispersion curves. Note that within Fig. \ref{ScatterPlots}
processes (i), (ii), and (iii) are identified by circular, square,
and triangular markers, respectively.

Backward intra-band scattering processes, which is the most widely
studied of Brillouin interactions, is commonly termed backward stimulated
Brillouin scattering \cite{boyd_nonlinear_2013,agrawal_nonlinear_2012};
references \cite{abedin_observation_2005,beugnot_brillouin_2014,behunin_long-lived_2015,beugnot_guided_2007,pant_-chip_2010}
are examples of this process. However, for historical reasons, the
terminology for forward intra-band and forward inter-band scattering
processes is somewhat more diverse. Thermally driven (or spontaneous)
forward intra-band scattering was first observed in optical fibers,
and identified as a noise process, under the name guided acoustic
wave Brillouin scattering (GAWBS) \cite{shelby_guided_1985}; references
\cite{zhong_depolarized_2015,elser_reduction_2006} are examples of
this spontaneous process. Stimulated forward intra-band scattering
processes have been described using the term (intra-modal) forward
stimulated Brillouin scattering \cite{renninger_guided-wave_2016,renninger_nonlinear_2016,renninger_forward_2016-1,shin_tailorable_2013,shin_control_2015,van_laer_interaction_2015,laer_net_2015,kittlaus_large_2016},
as well as using the more descriptive term stimulated Raman-like scattering
(SRLS) \cite{kang_all-optical_2010,butsch_optomechanical_2012}. 

Inter-band processes have also been observed through both spontaneous
and stimulated interactions under different names. Stimulated inter-band
coupling between co-propagating guided optical modes with different
polarization states has been termed stimulated inter-polarization
scattering (SIPS) \cite{kang_all-optical_2010}. In the context of
noise processes, the spontaneous version process has also been described
using the term de-polarized GAWBS or depolarization scattering \cite{zhong_depolarized_2015,elser_reduction_2006}.
Stimulated scattering between co-propagating guided optical modes
with distinct spatial distribution has also been described using the
term stimulated inter-modal scattering (SIMS) \cite{koehler_resolving_2016}
and stimulated inter-modal Brillouin scattering \cite{qiu_stimulated_2013}.

\bibliographystyle{naturemag}
\bibliography{ContOptomech3}

\end{document}